# Imaging Quasiparticle Interference in Bi$_2$Sr$_2$CaCu$_2$O$_{8+\delta}$

J. E. Hoffman[1], K. McElroy[1], D.-H. Lee[1], K. M Lang[1,2],

H. Eisaki[3,4,5], S. Uchida[5], J.C. Davis[1,6§]

[1]*Department of Physics, University of California, Berkeley, CA 94720-7300, USA.* [2]*Electromagnetic Technology Division, National Institute of Standards and Technology, Boulder, CO 80305-3328, USA.* [3]*Department of Applied Physics, Stanford University, Stanford, CA 94305-4090, USA.* [4]*Agency of Industrial Science and Technology, 1-1-1 Central 2, Umezono, Tsukuba, Ibaraki, 305-8568 Japan.* [5]*Department of Superconductivity, University of Tokyo, Tokyo, 113-8656 Japan.* [6]*Materials Sciences Division, Lawrence Berkeley National Laboratory, Berkeley, CA 94720, USA.* [§]*To whom correspondence should be addressed: email jcdavis@socrates.berkeley.edu.*

Scanning tunneling spectroscopy of the high-T$_c$ superconductor Bi$_2$Sr$_2$CaCu$_2$O$_{8+\delta}$ reveals weak, incommensurate, spatial modulations in the tunneling conductance. Images of these energy-dependent modulations are Fourier analyzed to yield the dispersion of their wavevectors. Comparison of the dispersions with photoemission spectroscopy data indicates that quasiparticle interference, due to elastic scattering between characteristic regions of momentum-space, provides a consistent explanation for the conductance modulations, without appeal to another order parameter. These results refocus attention on quasiparticle scattering processes as potential explanations for other incommensurate phenomena in the cuprates. The momentum-resolved tunneling spectroscopy demonstrated here also provides a new technique with which to study quasiparticles in correlated materials.

In an ideal metal, the Landau-quasiparticle eigenstates are Bloch wavefunctions characterized by wavevector $\vec{k}$ and energy $\varepsilon$. Their dispersion relation, $\varepsilon(\vec{k})$, can be measured with momentum resolved techniques such as angle resolved photoemission spectroscopy (ARPES). By contrast, real space imaging techniques, such as scanning tunneling microscopy (STM), cannot be used to measure $\varepsilon(\vec{k})$. This is because the local-density-of-states LDOS($E$) spectrum at a single location $\vec{r}$ is related to the $\vec{k}$-space eigenstates $\psi_k(\vec{r})$ by

$$\text{LDOS}(E,\vec{r}) \propto \sum_k |\psi_k(\vec{r})|^2 \, \delta(E - \varepsilon(\vec{k})) \qquad (1)$$

and substitution of a Bloch wavefunction into Eq. 1 shows LDOS($E$) to be spatially uniform.

When sources of disorder such as impurities or crystal defects are present, elastic scattering mixes eigenstates that have different $\vec{k}$ but are located on the same quasiparticle contour of constant energy (CCE) in $\vec{k}$-space. When scattering mixes states $\vec{k}_1$ and $\vec{k}_2$, an interference pattern with wavevector $\vec{q} = \vec{k}_2 - \vec{k}_1$ appears in the norm of the quasiparticle wavefunction and LDOS modulations with wavelength $\lambda = 2\pi/|\vec{q}|$ appear. These phenomena can be observed by STM as modulations of the differential tunneling conductance, which are often (imprecisely) referred to as "Friedel oscillations". STM studies of such conductance modulations have allowed the first direct probes of the quantum interference of electronic eigenstates in metals and semiconductors (*1-5*).

The Bogoliubov-quasiparticles in a Bardeen-Cooper-Schrieffer superconductor are also Bloch states but with dispersion

$$E_\pm(\vec{k}) = \pm\sqrt{\varepsilon(\vec{k})^2 + \Delta(\vec{k})^2} \qquad (2)$$

where $|\Delta(\vec{k})|$ is the $\vec{k}$-dependent magnitude of the energy gap at the Fermi surface (CCE for $e(\vec{k})=0$ in the normal state). Elastic scattering of Bogoliubov-quasiparticles can also result in conductance modulations. Quasiparticle standing waves have been imaged by STM in the conventional superconductor Nb (*6*) and in the CuO chains of YBCO (*7*). For the cuprates in general, it has long been proposed that conductance modulations due to quasiparticle scattering should occur, and that both the homogeneous electronic structure and superconducting gap anisotropy could be extracted from measurement of their properties (*8*).

When LDOS modulations are detected by STM in a given sample, certain CCE in $\vec{k}$-space can be reconstructed by analyzing the Fourier transform of the real-space LDOS(*E*) image (*3, 9, 10*). This is potentially a powerful technique because, for any *E*, it simultaneously yields real-space and momentum-space information on the wavefunctions, scattering processes, and dispersion of the quasiparticles. Application of this technique to the cuprate high temperature superconductors, as proposed in (*8*), would allow us to explore important issues including: (i) the physical processes dominating quasiparticle scattering, (ii) the degree to which the quasiparticles are well-defined and coherent, (iii) quasiparticle momentum-space structure and dispersion, and (iv) the relation between the commensurate and incommensurate magnetic signatures of the cuprates and quasiparticle scattering processes (*11-14*).

At low temperatures in $Bi_2Sr_2CaCu_2O_{8+\delta}$ (Bi-2212), a $\vec{k}$-dependent energy gap $\Delta(\vec{k})$ opens on the Fermi surface and new quasiparticles appear. Both the Fermi surface location in momentum-space, $\vec{k}_{FS}$, and its energy gap at these locations, $|\Delta(\vec{k}_{FS})|$, have been comprehensively studied by ARPES (*15-21*). Figure 1, A and B, shows representations of $\vec{k}_{FS}$ (black) and $|\Delta(\vec{k}_{FS})|$ (green). At the four gap-nodes, quasiparticle states exist down to zero energy while, at other $\vec{k}_{FS}$, a quasiparticle energy $E = \Delta(\vec{k}_{FS})$ is required to create the first excitation.

There is much evidence that quasiparticle scattering is important in Bi-2212. THz spectroscopy shows that low-temperature quasiparticle mean free paths in optimal Bi-2212 are about two orders of magnitude below that of optimal YBCO (*22*), indicating that appreciable quasiparticle scattering exists in Bi-2212. Furthermore, the strong nanoscale LDOS disorder observed between 25 and 65 meV by STM (*23-26*) is a potential source of this scattering.

We apply techniques of high-resolution LDOS imaging at 4.2 K on crystals grown by the floating zone method with superconducting transition temperature ($T_c$) ranging between underdoped ($T_c$ = 78 K) and slightly overdoped ($T_c$ = 85 K). The samples are cleaved at the BiO plane in cryogenic ultra high vacuum and immediately inserted into the STM head. Atomic resolution is achieved throughout the studies reported here. On these surfaces, we acquire maps of the differential tunneling conductance ($G = dI/dV$) measured at all locations ($x, y$) in the field of view (FOV). Because LDOS$(E = eV) \propto G(V)$, where $V$ is the sample bias voltage, this results in a two dimensional map of the LDOS at each energy ***E***.

We show in Fig. 2 a topographic image and three LDOS maps for quasiparticle energies centered at 12, 16, and 22 meV, all acquired in the same 650 Å FOV with 1.3 Å spatial resolution. Periodic LDOS modulations are evident in all images (although one also sees remnants of impurity scattering at low energies and of gap disorder at high energies). Notably, quite different spatial patterns and wavelengths are observed at each energy.

To explore the evolution of these LDOS modulations with energy, we take the amplitude of the Fourier transform, FT($\vec{E}, \vec{q}$), of LDOS maps measured in a 600 Å FOV (Fig. 3). We also show the simultaneously acquired topographic image and the relative orientation of the CuO$_2$ reciprocal unit cell and $\Delta(\vec{k}_{FS})$. Although the regions of $\vec{q}$-space displaying LDOS modulation intensity are changing rapidly and intricately with energy, two important effects can be discerned. First, local peaks in FT($\vec{E}, \vec{q}$), whose $\vec{q}$-vectors are oriented toward the ($\pm\pi$, 0) and (0, $\pm\pi$) directions, appear at finite $|\vec{q}|$ at very low energy and then move steadily inwards towards (0, 0) (i.e., $|\vec{q}|$ decreases as $E$ increases). Second, peaks in the FT($E,\vec{q}$), with $\vec{q}$-vectors along the (±π, ±π) directions, appear and move steadily to larger $|\vec{q}|$ with increasing energy. The same set of phenomena have been observed for all six samples we have studied in this manner, but the exact dispersion of these peaks varies systematically with doping.

In Fig. 4, A and B, we plot the measured value of FT($E,\vec{q}$) from Fig. 3 versus $|\vec{q}|$ along the (π, 0) and (π, π) directions, at several energies. The locations of the FT($E,\vec{q}$) peaks are measured by fitting an exponential decay plus a Lorentzian. The dispersion in this unprocessed data is obvious. In Fig. 4, C and D, we show the energy dependence of the peaks in FT($E,\vec{q}$) for $\vec{q}$-vectors oriented toward the (π, 0) direction [$\vec{q}_A(E)$], and towards the (π, π) direction [$\vec{q}_B(E)$]. The dispersions of these two types of conductance modulations were analyzed in detail for data for three of the samples. One is underdoped with mean energy gap value $\overline{\Delta}$ = 50.2 meV (red squares), the second is near optimal with $\overline{\Delta}$ = 43.7 meV (green circles), and the third is slightly overdoped with $\overline{\Delta}$ = 36.7 meV (blue triangles). In general, one can see that at fixed $E$, the $\vec{q}_A(E)$ become shorter, whereas the $\vec{q}_B(E)$ become longer, as the doping is increased.

Which scattering processes could be responsible for the phenomena in Figs. 2 to 4? In general, a particular wavevector $\vec{q}$ can dominate the quasiparticle interference at energy $E$, if the $\vec{k}$-pairs on the CCE connected by $\vec{q}$ have a large joint density of states (DOS). A full theoretical treatment of this issue is beyond the scope of this paper. Instead we introduce a simplified picture that captures many of the key elements. Equation 2 shows how, due to the $\vec{k}$ dependence of the gap magnitude $|\Delta(\vec{k})|$, different $\vec{k}_{FS}$ values have a different minimum $E$ at which quasiparticles may appear. From the fourfold symmetry, this means that at any non-zero energy $E$ there are only eight possible $\vec{k}_{FS}$ values at which $|\Delta(\vec{k}_{FS})|$ = $E$ in the first Brillouin zone. The highest joint-DOS for quasiparticle scattering at this energy occurs at $\vec{q}$-vectors connecting these eight points (Fig. 1B) (27). One might expect the interference-induced conductance modulations to occur at some of these $\vec{q}$-vectors.

In this context we consider data from the (π, 0) direction. We denote the vector that connects points with the same Δ but on opposite almost parallel Fermi surface segments as $\vec{q}_A$ (blue arrows in Fig. 1). These wavevectors are parallel to (0, ±π) or (±π, 0) but have different $|\vec{q}_A|$ depending on the Δ at the points being connected. We estimate the expected $\vec{q}_A(\Delta)$ in our model using ARPES measurements of $|\Delta(\vec{k})|$ and locations of $\vec{k}_{FS}$ (17). The result is shown as grey bands in Fig. 4C. The ARPES-derived results and our measured $\vec{q}_A(E)$ are in excellent quantitative agreement. With increasing doping, the measured range of $\vec{q}_A$ becomes systematically shorter (Fig. 4C). This would be expected if increased hole density expands the hole pocket and moves the almost-parallel sections of the Fermi surface closer together (Fig. 1B).

The (π, π)-oriented peaks in the FT($E, \vec{q}$) evolve very differently with energy than those oriented toward (π, 0). Their dispersion has opposite sign and is substantially slower. We consider scattering that connects the same range of $\vec{k}$-states on the Fermi surface as for the (π, 0)-oriented process, but now diagonally across the inside of the hole pocket by the vector $\vec{q}_B$ parallel to (π, π) (Fig. 1, red arrows). We can again estimate the expected $|\vec{q}_B(\Delta)|$ using ARPES data (17). The result is shown as a grey band in Fig. 4D. The ARPES-derived results and our measured $\vec{q}_B(E)$ are in excellent quantitative agreement. Furthermore, with increasing doping, the range of $\vec{q}_B(\Delta)$ moves to higher values, again as expected if increased hole density increases the area of the hole pocket and the distance between relevant sections of the Fermi surface (Fig. 1B).

These results are relevant to several issues. First, reported $4a_0$ periodic LDOS modulations at $E = 25$ meV in Bi-2212 (28) have been interpreted as stemming from the simultaneous existence of another electronic ordered-state. However, one of us (DHL) proposed that instead, such conductance modulations are quasiparticle interference effects (*27*). The data in Fig. 2, B to D, appear to be due primarily to this latter effect because (i) all modulations have appreciable dispersion, (ii) the dispersions are consistent with scattering between the identified $\vec{k}$-space regions of high joint-DOS, and (iii) the evolution of the dispersions with doping is consistent with expected changes in the Fermi surface. Thus, it appears that quasiparticle band-structure effects play the primary role and must be understood before departures from them can be ascribed to other order parameters.

A second issue is the unknown relation between the relatively strong "checkerboard" modulations around the vortex core (*29*) and the weak modulations in zero-field discussed here and in (*28*). The field-induced LDOS-"checkerboard" is localized in a small region around the core, centered on $\boldsymbol{E}$ = 7 meV, has a spatial wavelength of ~4.2 ± 0.4$a_0$ , and is between 10 and 100 times as intense as the modulations near E = 7 meV (with a different wavevector) discussed in this study. There is a possibility that the field-induced "checkerboard" could be produced by quasiparticle scattering off the vortex core, or it could be an indirect signature of a second order parameter stabilized near the core, or it could even be some combination of the two. New experiments will be required to distinguish between these cases.

A third issue is the range of coherent quasiparticle interference patterns and the imprecise location of their scattering centers. The smallest $\vec{q}$-space extent of an FT($\boldsymbol{E},\vec{q}$) peak in Fig. 3 is $\Delta q$ ~ 0.1π/$a_0$. This indicates that the longest coherence length for any modulation is $\ell$ ~ 80 Å. Several phenomena such as gap disorder [with patch size ~30 Å (*23-26*)], impurity resonances (with spacing ~100 Å visible in Fig. 2B), or oxygen atoms (with spacing ~13 Å at this nominal doping) may influence this coherence length. We cannot at present identify which factors are key.

The fourth, and possibly most important, point is that quasiparticle scattering between high joint-DOS regions of $\vec{k}$-space has now received direct experimental support as a mechanism for incommensurate, dispersive, spatial modulations of the superconducting electronic structure. A related process, in which a quasiparticle is scattered across the Fermi energy into a quasi-hole and vice versa, has been theoretically discussed as a potential explanation (*11-14*) for the incommensurate, dispersive, magnetic phenomena detected by neutron scattering in the cuprates (*30*). Renewed exploration of such a scattering-related explanation for these phenomena may therefore be appropriate.

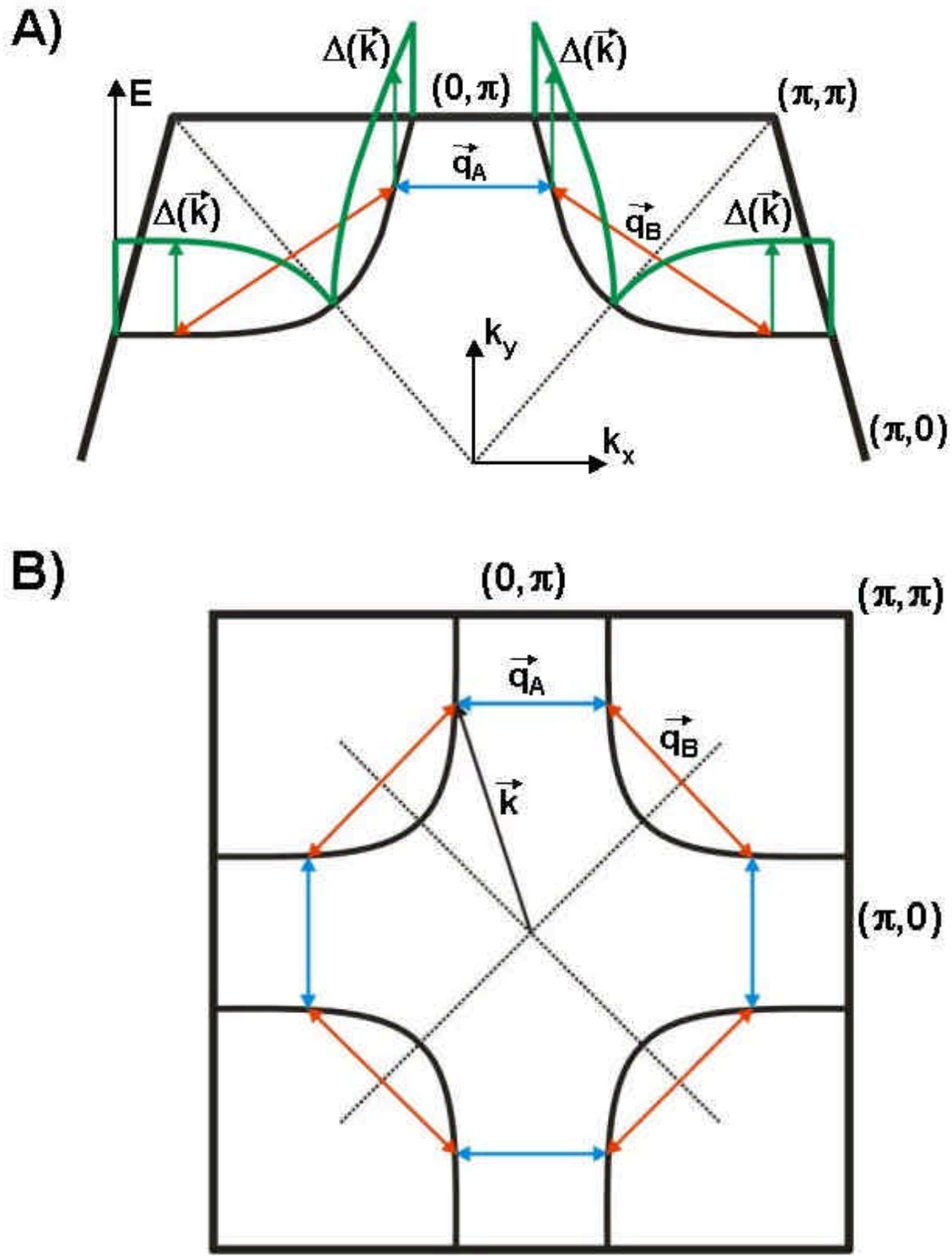

**Fig. 1.** **(A)** A perspective view of the superconducting energy gap Δ (green) as a function of the location along the Fermi surface (black). **(B)** Schematic Fermi surface of Bi-2212. Vectors connecting the eight areas of the Fermi surface with identical |Δ| are shown and labeled by blue and red arrows depending on the type of elastic scattering process at $E = |\Delta|$ that connects them.

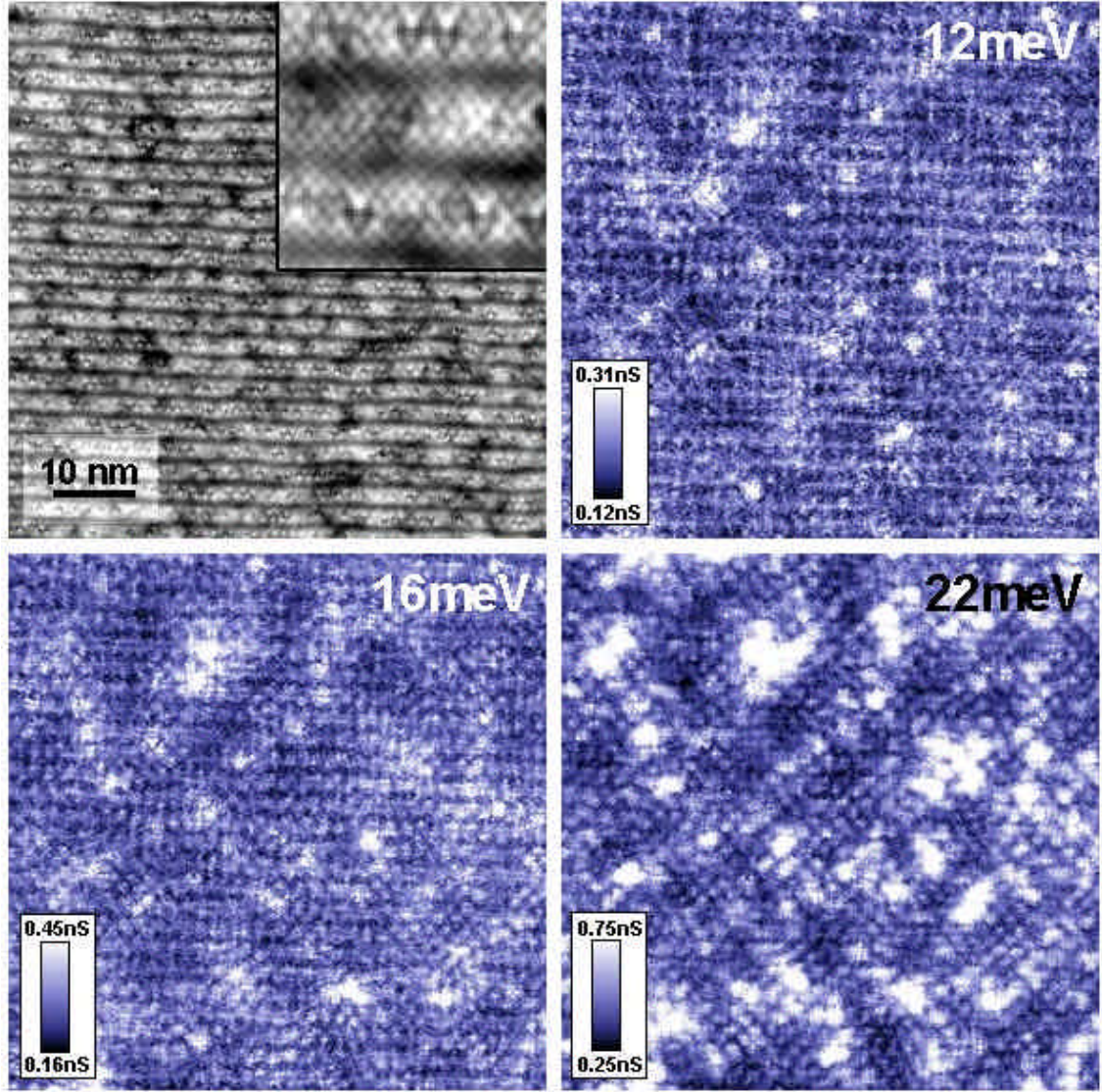

**Fig. 2.** Four unprocessed images of a single 650 Å field of view with 1.3 Å resolution: **(A)** Topography (inset magnification: x2), **(B)** 12 meV LDOS, **(C)** 16 meV LDOS, and **(D)** 22 meV LDOS. All measurements reported in this paper were obtained at a junction resistance of 1 GΩ set at a bias voltage of -100 mV, so the total junction conductance is $10^3$ picoSiemens (pS). All LDOS images were acquired with a bias modulation amplitude of 2 mV root mean square. From the atomic resolution in (A) one can see the direction of the Cu-O bonds and the incommensurate supermodulation of wavelength ~26 Å at 45° to the bonds. The LDOS images show checkerboard-like modulations of the LDOS: (B) 45° to the Cu-O bonds at $\vec{q}_B \sim 0.21(\pm 2\pi/a_0, \pm 2\pi/a_0)$; (C) both 45° to and along the Cu-O bonds at $\vec{q}_B \sim 0.25(\pm 2\pi/a_0, \pm 2\pi/a_0)$ but also at $\vec{q}_A \sim 0.22(0, \pm 2\pi/a_0)$ and $\vec{q}_A \sim 0.22(\pm 2\pi/a_0, 0)$; (D) along the Cu-O bonds at $\vec{q}_A \sim 0.20(0, \pm 2\pi/a_0)$ and $\vec{q}_A \sim 0.20(\pm 2\pi/a_0, 0)$.

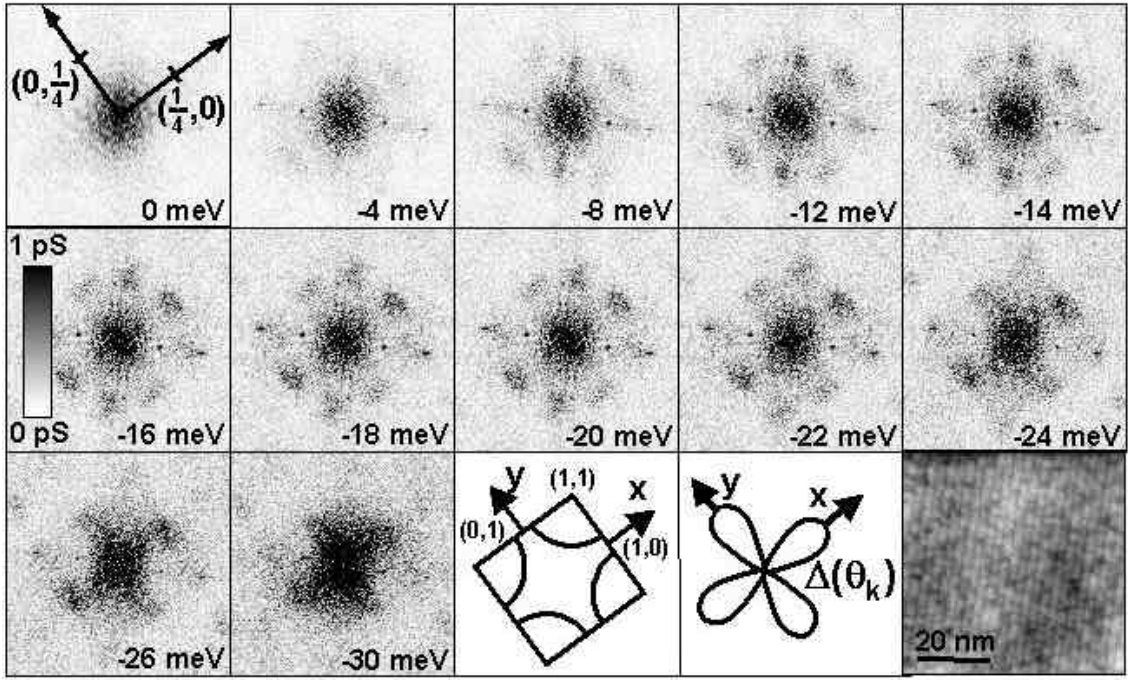

**Fig. 3.** A series of 12 Fourier transforms of LDOS images measured on a 600 Å square FOV at the energies shown in each panel. The origin and points (1/4, 0) $2\pi/a_0$ and (0, 1/4) $2\pi/a_0$ are labeled. The local maxima in these images represent the dominant $\vec{q}$ associated with quasiparticle scattering at each energy $E$. The scattering wavevectors oriented parallel to $(\pm\pi, 0)$ and $(0, \pm\pi)$ become shorter as energy rises. In contrast, those oriented parallel to $(\pm\pi, \pm\pi)$ become longer as energy rises. The harmonics of the supermodulation are sharp features observed at the same location in all FT($E, \vec{q}$) images. The simultaneously acquired topographic image plus the orientation of the Fermi surface and $\Delta(\vec{k})$ are shown.

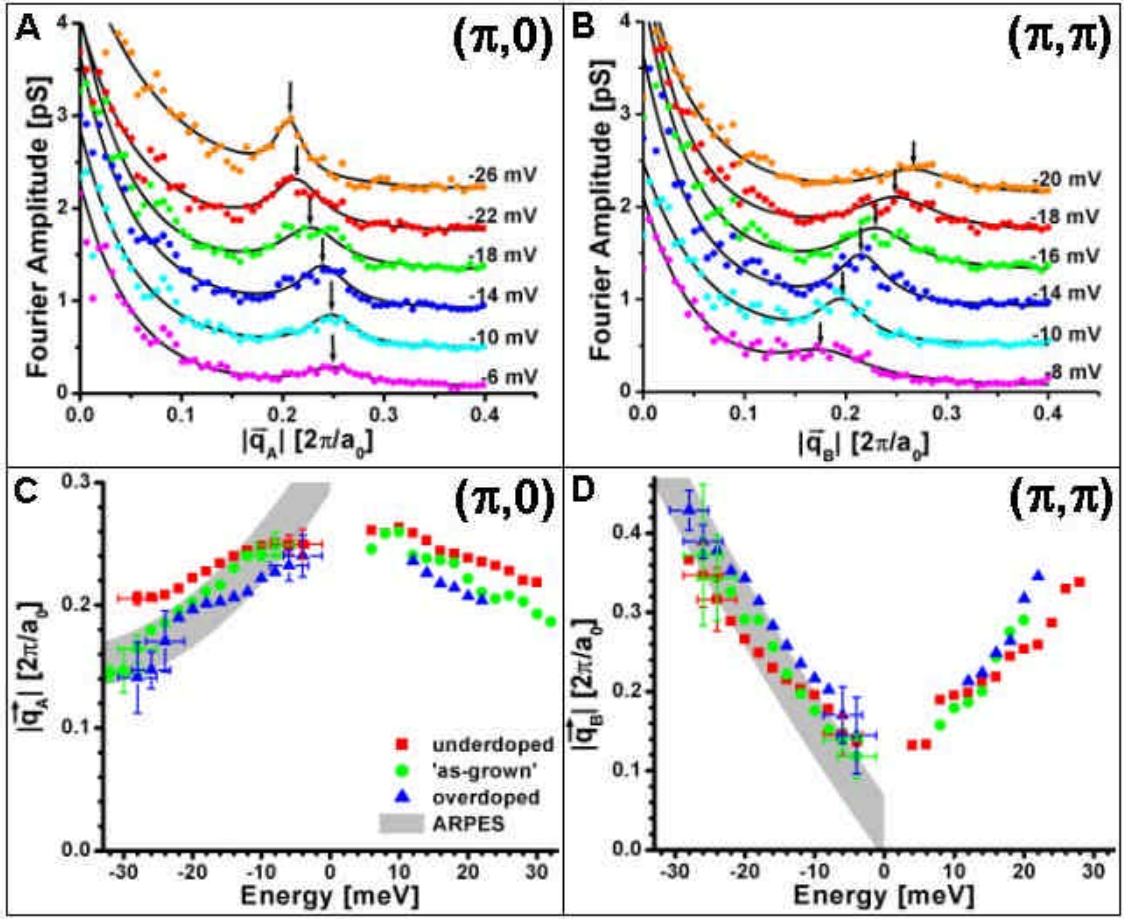

**Fig. 4.** The amplitude of FT($E, \vec{q}$) versus $|\vec{q}|$ along **(A)** the ($\pi$, 0) direction and **(B)** the ($\pi$, $\pi$) direction are shown for six quasiparticle energies. The data are shifted vertically relative to each other by 0.4 pS for clarity. Each solid black line is a fit to an exponential decay from $\vec{q} = 0$ plus a Lorentzian peak. An arrow points to the local maximum in $|\vec{q}|$ for each energy. The measured $\vec{q}_A(E)$ for **(C)** the ($\pi$, 0) direction and **(D)** the ($\pi$, $\pi$) direction are plotted for samples of three different dopings. The error bars shown at the end of the data ranges are representative of the uncertainty in identifying the location due to the peak width. The shaded grey bands represent the expected dependence $\vec{q}(E)$ derived from ARPES using the model described in the text. Note that ARPES measures only the electron-like portion of the spectrum. The width of the band represents uncertainties in the published ARPES data. These ARPES data from a $T_c$ = 87 K Bi-2212 sample (*17*) should be compared with data from our similarly doped sample (green circles), which are in excellent agreement with the simplified model.